\documentclass[11pt]{article}%
\usepackage{amsfonts}
\usepackage{amsmath}
\usepackage{amssymb}
\usepackage{graphicx}%
\providecommand{\U}[1]{\protect\rule{.1in}{.1in}}
\newtheorem{theorem}{Theorem}
\newtheorem{acknowledgement}[theorem]{Acknowledgement}

\newtheorem{lemma}[theorem]{Lemma}

\newtheorem{proposition}[theorem]{Proposition}

\newenvironment{proof}[1][Proof]{\noindent\textbf{#1.} }{\ \rule{0.5em}{0.5em}}
\begin{document}

\title{Imprints of the Quantum World in Classical Mechanics}
\author{Maurice A. de Gosson\thanks{Maurice de Gosson has been financed by the
Austrian Research Agency FWF (Projekt \textquotedblleft Symplectic Geometry
and Applications to TFA and QM\textquotedblright, Projektnummer P20442-N13).}\\\ \textit{Universit\"{a}t Wien, NuHAG}\\\textit{Fakult\"{a}t f\"{u}r Mathematik }\\\textit{A-1090 Wien}
\and Basil Hiley\\\ \textit{TPRU, Birkbeck}\\\textit{University of London}\\\textit{London, WC1E 7HX}}
\maketitle

\begin{abstract}
The imprints left by quantum mechanics in classical (Hamiltonian) mechanics
are much more numerous than is usually believed. We show that the
Schr\"{o}dinger equation for a nonrelativistic spinless particle is a
classical equation which is equivalent to Hamilton's equations. Our discussion
is quite general, and incorporates time-dependent systems. This gives us the
opportunity of discussing the group of Hamiltonian canonical transformations
which is a non-linear variant of the usual symplectic group.

\end{abstract}

\section{Introduction}

\begin{quotation}
\textit{\textquotedblleft Where did that [the Schr\"{o}dinger equation] come
from? Nowhere. It came out of the mind of Schr\"{o}dinger, invented in his
struggle to find an understanding of the experimental observations in the real
world. \textquotedblright\ (Richard Feynman in \cite{Feynman}.)}
\end{quotation}

Similar statements abound in the physical literature; they are found in both
introductory and advanced text on quantum mechanics, and we can read them on
the web in various blogs and forums. However, they are strictly speaking
\emph{not true}; already in 1966 Nelson \cite{Nelson} showed that
Schr\"{o}dinger's equation could be derived from Newtonian mechanics; he
however introduced some extra physical assumptions (stochasticity); also see
the more recent paper by Hall and Reginatto \cite{Hall}, who use the
uncertainty principle. We also mention, because of its historical interest,
Feynman's non-rigorous approach in \cite{Feynman2} (see Derbes' comments
\cite{Derbes} of Feynman's paper).

In the present paper we will show that one can \emph{mathematically} derive
\emph{rigorously} from Hamiltonian mechanics, the Schr\"{o}dinger equation%
\[
i\hbar\frac{\partial\psi}{\partial t}=H(x,-\hbar\nabla_{x},t)\psi
\]
and this without any recourse to any physical `quantum-mechanical' hypotheses.
In fact we will show that there is a surprising one-to-one and onto
correspondence between Hamiltonian flows and the quantum evolution group,
which only becomes apparent if one uses a deep property of symplectic
covariance together with Stone's theorem on one-parameter groups of unitary
operators. Schr\"{o}dinger \cite{es26a} was led to his equation from his
knowledge of the classical Hamilton-Jacobi approach which has a close
connection with the eikonal of classical wave theory. His original proposal
was not firmly based on rigorous mathematics as he himself acknowledged but
his intuition was correct. Attempts to provide a detailed relationship between
classical and quantum mechanics has remained somewhat of a problem because we
have been left with two widely different mathematical formulations, one
involving particles evolving under dynamical laws in a phase space, the other
involving operators and waves operating in a Hilbert space. This leaves the
impression that there are two very different worlds, the classical and the
quantum. However it is clear that we do not inhabit two different worlds and
so we are left with the puzzle as how to relate the two formalisms?

Already in studying the properties of light, we see in one phenomena both
aspects. Geometric optics gives us light rays travelling in straight lines,
while wave optics gives us interference and diffraction. In this phenomena we
see the essential mathematics emerging. Ray optics emerges from symplectic
geometry, while wave optics emerges from the geometry of the covering group of
the symplectic group, $\operatorname*{Sp}(2n,\mathbb{R})$, namely the
metaplectic group, $\operatorname*{Mp}(2n,\mathbb{R})$. Ray tracing involves
symplectic flows, while the wave evolution is identified with the metaplectic
flows. These are not separate flows but deeply related as the metaplectic
flows can be mathematically `lifted' from symplectic flows and the metaplectic
flows can be projected onto symplectic flows.

The fact that classical mechanics has the same mathematical structure as ray
optics, namely the symplectic geometry, suggest that quantum mechanics in its
wave description could be related to such a covering structure. Indeed if we
restrict ourselves to quadratic Hamiltonians, we find that the metaplectic
flow is determined exactly by Schr\"{o}dinger's equation, in which the
classical Hamiltonian is replaced by the Weyl Hamiltonian,
$H\overset{\text{Weyl}}{\leftrightarrow}\widehat{H}$. Unfortunately the
generalisation to all physically relevant Hamiltonians has always floundered
on the mathematical difficulties involved in investigating the covering
properties of general group of canonical transformations which involve
non-linear transformations. However such difficulties should not force us into
a `two world' situation.

Indeed we have shown recently \cite{FP,PR} that Hamiltonian mechanics, in its
symplectic formulation, in a sense reproduces what is considered as being one
of the hallmarks of quantum mechanics, the uncertainty principle in its strong
form (the Schr\"{o}dinger--Robertson inequalities); the argument is based on a
deep and new property of canonical transformations, Gromov's \cite{Gromov}
non-squeezing theorem, alias \textquotedblleft the principle of the symplectic
camel\textquotedblright.

Up to some technical assumptions on the Hamiltonian function ensuring us that
the flows they generate exist for all times, we will in fact show that:

\emph{There is a one-to-one and onto correspondence between Hamiltonian flows
generated by a Hamiltonian $H$ and strongly continuous unitary one-parameter
groups satisfying Schr\"{o}dinger's equation with Hamiltonian operator
$\widehat{H}=H(x,-i\hbar\nabla_{x},t)$ obtained from $H$ by Weyl quantization.
Equivalently, the Hamilton equations%
\[
\dot{x}=\nabla_{p}H(x,p,t)\ ,\ \dot{p}=-\nabla_{x}H(x,p,t)
\]
are mathematically rigorously equivalent to Schr\"{o}dinger's equation}%
\[
i\hbar\frac{\partial\psi}{\partial t}=H(x,-i\hbar\nabla_{x},t)\psi.
\]

This property is actually well-known (at least in mathematics \cite{Birk,GS})
for the linear flows arising from quadratic Hamiltonian functions; the new
result is that this surprising correspondence between Hamiltonian and quantum
flows has a quite general nature.

The proof relies on two deep mathematical results

\begin{itemize}
\item The first is that the Weyl correspondence is the \emph{only
}quantization\emph{ (that is, }pseudodifferential calculus) which is
symplectically covariant. We emphasize the word \textquotedblleft
only\textquotedblright, because while it is reasonably well-known, both in
mathematics and physics that Weyl operators are symplectically covariant, the
converse, namely the only operators that are symplectically covariant are Weyl
operators, is largely ignored in the literature; this uniqueness is the key to
our argument.

\item The second is Stone's theorem from functional analysis which concerns
the infinitesimal generators of strongly continuous unitary one-parameter
groups. It states that each such one-parameter group determines a self-adjoint operator.
\end{itemize}

In Section \ref{secham} we will define and study the group $\operatorname{Ham}%
(2n,\mathbb{R})$ of all Hamiltonian canonical transformations following
Banyaga \cite{Banyaga} and Polterovich \cite{Polter}; the systematic
consideration of general time-dependent Hamiltonians will make our task
considerably easier than if we had limited ourselves to the case where $H$
only depends on the phase space variables, $z=(x,p)$.

The hypothesis of a bounded phase space (implying a fortiori a bounded
Universe) is not essential; it is added as an ad hoc assumption in order to
avoid mathematical difficulties related to self-adjointness and the fact that
Hamiltonian vector fields need not be complete.

There are related issues that will not be addressed in this paper. For
instance, one could envisage extending the results of this paper to the
Schr\"{o}dinger equation\ on Riemannian manifolds (compact or not). Since our
main result links Hamiltonian flows with quantum flows, one might wonder if
the control of the former allows to derive some control of the other. This is
a question that certainly deserves to investigated, but we will leave it to
forthcoming publication. It would be also interesting to discuss recent
results of Schmelzer \cite{schmelz1,schmelz2} about foundational issues in our
context.\bigskip

\textbf{CAVEAT\ LECTOR:} We are \emph{not} claiming that we are deriving
quantum mechanics from classical mechanics; what we are doing is the
following: knowing that quantum mechanics exists, we show that the
mathematical formulation of quantum mechanics in its Schr\"{o}dinger
formulation lies within Hamiltonian mechanics. This does not imply that
quantum mechanics --as a \emph{physical} theory-- can be reduced to classical
mechanics. For as Mackey (who used to see quantum mechanics as a
\emph{refinement} of Hamiltonian mechanics) stresses in \cite{Mackey} (p.
106), quantum mechanics is not just an algorithm for attaching a self-adjoint
operator to every classical Hamiltonian, because such a program would overlook
many facts: first, quantum mechanics rules out a large number of conceivable
Hamiltonians, and secondly there are features of quantum mechanics (such as
spin) which do not manifest themselves in the classical limit. However, adds
Mackey, one can argue convincingly that classical mechanics must have
independent existence since it is only what quantum mechanics looks like under
certain limiting conditions. Our program does not contradict Mackey's insight
because our approach shows that the imprint left by quantum mechanics in
classical mechanics is more than could be expected at first sight from the
limiting conditions.

\section{The Dynamical Groups $\operatorname*{Sp}(2n,\mathbb{R)}%
,\operatorname*{Mp}(2n,\mathbb{R})$ and $Ham(2n,\mathbb{R)}$}

\subsection{Motivation from Optics}

We recall that ray tracing in Gaussian optics uses an identical formulation of
classical mechanics, which for the purposes of this paper can be written in
the form
\[
z_{t}=f_{t}^{H}(z_{0})
\]
where the phase variables $z=(x_{1}\dots x_{n},p_{1}\dots p_{n})$ and
$f_{t}^{H}$ is the flow determined by Hamilton's equations of motion
\begin{equation}
\dot{x}=\nabla_{p}H(x,p)\text{ \ , \ }\dot{p}=-\nabla_{x}H(x,p). \label{HEQ}%
\end{equation}
where $H(x,p)$ is some physically relevant Hamiltonian.

We now ask the question, \textquotedblleft Is it possible to lift this flow
onto the covering group of $\operatorname*{Sp}(2n,\mathbb{R)}$ so as to
display possible wave properties of the symplectic flow?" The answer is,
\textquotedblleft Yes, provided we restrict ourselves to quadratic
Hamiltonians.\textquotedblright

To see how this works let us write the quadratic Hamiltonian in the form
\begin{equation}
H(z,t)=\frac{1}{2}z^{T}M(t)z=\frac{1}{2}%
\begin{pmatrix}
x\\
p
\end{pmatrix}
^{T}M(t)%
\begin{pmatrix}
x\\
p
\end{pmatrix}
\label{H}%
\end{equation}
where $M(t)$ is a $2n\times2n$ symmetric matrix depending in a $C^{\infty}$
fashion on the parameter $t$. Let us now shortly describe the metaplectic
representation of the symplectic group $\operatorname*{Sp}(2n,\mathbb{R)}$.
Let $W$ be a real quadratic form of the type%
\begin{equation}
W(x,x^{\prime})=\tfrac{1}{2}Px\cdot x-Lx\cdot x^{\prime}+\tfrac{1}%
{2}Qx^{\prime}\cdot x^{\prime} \label{W}%
\end{equation}
with $P=P^{T}$, $Q=Q^{T}$, and $\det L\neq0$; we are writing $Px\cdot x$ for
$x^{T}Px$, etc. The function $W$ has a very precise meaning in optics, where
it is called the eikonal (see Chapter 1 in Guillemin and Sternberg's book
\cite{GS} for a discussion of the relation between physical optics and the
metaplectic group); it is also well-known in Hamiltonian mechanics, where it
is called a generating function, or Hamilton's two-point function and is
closely related to action (see de Gosson \cite{ICP,Birk} and the references
therein). To such a function $W$ one associates a linear canonical
transformation $s^{W}$ by the formula%
\begin{equation}
(x,p)=s^{W}(x^{\prime},p^{\prime}); \label{fw}%
\end{equation}
explicitly $s^{W}$ is then identified with the symplectic matrix%
\begin{equation}%
\begin{pmatrix}
L^{-1}Q & L^{-1}\\
PL^{-1}Q-L^{T} & L^{-1}P
\end{pmatrix}
. \label{splq}%
\end{equation}
One shows (de Gosson \cite{ICP,Birk}) that every symplectic matrix $s$ can be
written (non-uniquely) as a product of two such matrices: $s=s^{W}%
s^{W^{\prime}}$ hence the $s^{W}$ generate $\operatorname*{Sp}(2n,\mathbb{R)}%
$. Denoting by $m\pi$ the choice of an argument for $\det L$ modulo $2\pi$ we
define the Fourier quadratic transformation $S^{W}$ by%

\begin{equation}
S^{W}\psi(x)=\left(  \frac{1}{2\pi i\hbar}\right)  ^{n/2}i^{m}\sqrt{|\det
L|}\int_{\mathbb{R}^{n}}e^{\frac{i}{\hbar}W(x,x^{\prime})}\psi(x^{\prime
})dx^{\prime}\label{swmh}%
\end{equation}
where $m$ is the Maslov index (see e.g. de Gosson \cite{Birk}, Chapter 7). One
easily checks that the operators $S^{W}$ are unitary isometries of
$L^{2}(\mathbb{R}^{n})$ and that the inverse of $S^{W}$ is the operator
$S^{W^{\ast}}$ where $W^{\ast}(x,x^{\prime})=-W(x^{\prime},x)$ (it amounts
replacing the triple $(P,L,Q)$ in (\ref{W}) by $(-Q,-L^{T},-P)$); notice that
the choice $W(x,x^{\prime})=x\cdot x^{\prime}$ leads to the usual Fourier
transform up to the factor $i^{-n/2}$ and that we then have $s^{W}=J$ (the
standard symplectic matrix). The operators $S^{W}$ thus generate a group of
unitary operators acting on the square-integrable functions; this group is
precisely the metaplectic group $\operatorname*{Mp}(2n,\mathbb{R)}$. It is
closed related to the symplectic group; one shows that it is a true unitary
representation of the double covering $\operatorname{Sp}_{2}(2n,\mathbb{R})$
of $\operatorname*{Sp}(2n,\mathbb{R)}$, and that the mapping $\Pi
^{\operatorname*{Mp}}:\operatorname*{Mp}(2n,\mathbb{R)\longrightarrow
}\operatorname{Sp}(2n,\mathbb{R})$ defined by $\Pi^{\operatorname*{Mp}}%
(F^{W})=f^{W}$ where $f^{W}$ is defined by (\ref{fw}) is two-to-one. The
homomorphism $\operatorname{Sp}_{2}(2n,\mathbb{R})\thickapprox
\operatorname{Mp}(2n,\mathbb{R})$ allows us to associate to the curve
$t\longmapsto f_{t}$ in $\operatorname{Sp}(2n,\mathbb{R})$ a unique curve
$t\longmapsto F_{t}$ in $\operatorname{Mp}(2n,\mathbb{R})$ such that $F_{0}$
is the identity operator. This means in terms of the projection that the
relation $\Pi^{\operatorname*{Mp}}(F_{t}^{W})=f_{t}^{H}$ unambiguously define
a one-to-one correspondence between continuous curves in $\operatorname*{Sp}%
(2n,\mathbb{R)}$ passing through the identity and continuous curves in
$\operatorname*{Mp}(2n,\mathbb{R)}$ passing through the identity.

Now choose a function $\psi_{0}$ in some subspace of infinitely differentiable
functions in $L^{2}(\mathbb{R}^{n})$ and set $\psi_{t}=F_{t}\psi_{0}$ where
$(F_{t})$ is the curve in $\operatorname*{Mp}(2n,\mathbb{R)}$ associated by
the procedure outlined above to a Hamilton flow $(f_{t})$ determined by a
quadratic Hamiltonian function We have $\psi_{t}\in L^{2}(\mathbb{R}^{n})$ for
each $t$ and the function $\psi_{t}$ satisfies the Schr\"{o}dinger-like
equation
\begin{equation}
i\frac{\partial\psi_{t}}{\partial t}=\widehat{H}\psi_{t} \label{S1}%
\end{equation}
where $\widehat{H}$ is a partial differential operator obtained from $H$ using
the Weyl ordering. That is, we perform the substitution $p\longrightarrow
-i\nabla_{x}$ in the Hamiltonian (\ref{H}) after having written it in the
form
\[
H(z,t)=\frac{1}{2}A(t)x\cdot x+B(t)p\cdot x+B^{T}(t)x\cdot p+\frac{1}%
{2}C(t)p\cdot p
\]
where $M(t)=%
\begin{pmatrix}
A(t) & B(t)\\
B^{T}(t) & C(t)
\end{pmatrix}
$ with $A(t)=A^{T}(t)$ and $C(t)=C^{T}(t)$. Using the obvious identity
$B\nabla_{x}\cdot x=x\cdot B\nabla_{x}+i\operatorname{Tr}B$ we thus obtain
\[
\widehat{H}=\frac{1}{2}A(t)x\cdot x-ix\cdot B(t)\nabla_{x}+\frac{1}%
{2}C(t)\nabla_{x}\cdot\nabla_{x}-\frac{i}{2}\operatorname{Tr}B(t)
\]
from which it is easy to verify that the operator $\widehat{H}$ is symmetric
(in fact self-adjoint) for the $L^{2}$ scalar product. One formally writes
\begin{equation}
\widehat{H}=H(x,-i\hbar\nabla_{x},t). \label{hat}%
\end{equation}
which is obtained from $H$ by the substitution $p_{j}\longrightarrow
-i\hbar\partial/\partial x_{j}$; it is of course \emph{formally} the usual
\textquotedblleft quantization\textquotedblright\ of the Hamiltonian function
$H$ of standard quantum mechanics.

It looks as if we have gained nothing by the process we have gone through, but
this conclusion is not correct. We have shown that there is a deep relation
between classical and quantum mechanics in this instance. It is the
isomorphism $\operatorname{Sp}_{2}(2n,\mathbb{R})\approx\operatorname*{Mp}%
(2n,\mathbb{R})$ that allows us to associate the classical flow $f_{t}^{H}$
with the quantum flow $F_{t}^{W}$. In this sense we have been able to derive
the Schr\"{o}dinger equation from the classical equations of motion.

\subsection{The Group $\operatorname{Ham}(2n,\mathbb{R})$\label{secham}}

The group $\operatorname{Ham}(2n,\mathbb{R})$ of Hamiltonian canonical
transformations is the non-linear analogue of the symplectic group
$\operatorname{Sp}(2n,\mathbb{R})$. What we would be looking for if we are to
generalise the quadratic Hamiltonian result is a unitary representation of
some covering structure of Ham($2n$). This possibility has not been explored
in the physical literature (to the best of the knowledge of the present
authors). It is the purpose of this paper to show how these results can be
extended to all physically relevant Hamiltonians. To this aim, the next two
sections discuss the symplectic covariance of Hamiltonian mechanics and the
corresponding symplectic covariance of Weyl quantisation. It is through these
two notions that we can show how the Schr\"{o}dinger equation is directly
obtained from classical Hamiltonian mechanics.

\section{Symplectic covariance of Hamiltonian mechanics}

Let $H$ now be an arbitrary Hamiltonian function, that is, a function
$H(x,p,t)$ which we assume to be at least once continuously differentiable in
the variables $x_{1},...,x_{n};p_{1},...,p_{n}$ and $t$. The flow $(f_{t}%
^{H})$ determined by these equations (one also says \textquotedblleft
generated by $H$\textquotedblright) consists of the mappings $f_{t}%
^{H}:\mathbb{R}^{2n}\longrightarrow\mathbb{R}^{2n}$ which associate to a point
$z=(x,p)$ at initial time $t=0$ the value $z_{t}=(x_{t},p_{t})$ of the
solutions at time $t$ of the differential equations
\begin{equation}
\dot{x}=\nabla_{p}H(x,p,t)\text{ \ , \ }\dot{p}=-\nabla_{x}H(x,p,t).
\label{HEQ3}%
\end{equation}
Thus, by definition, $z_{t}=f_{t}^{H}(z_{0})$. Generically the flow is only
defined in a neighborhood of $t=0$; we will later see that the consideration
of compactly supported Hamiltonian functions eliminates these difficulties.

One of the problems we face is that for time dependent Hamiltonians, the
corresponding flows $f_{t}^{H(t)}$ are such that
\[
f_{t}^{H(t)}f_{t}^{H(t^{\prime})}\neq f_{t+t^{\prime}}^{H(t+t^{\prime})}%
\]
and therefore does no have the required group property. In subsection 2.3 we
will show how a group Ham($2n,\mathbb{R)}$ can be constructed. In the mean
time we note that when $H$ does not dependent on time: $\partial H/\partial
t=0$, the system of differential equations (\ref{HEQ3}) is autonomous and the
flow $(f_{t}^{H})$ satisfies the one-parameter group property $f_{t}%
^{H}f_{t^{\prime}}^{H}=f_{t+t^{\prime}}^{H}$ whenever the mappings $f_{t}^{H}%
$, $f_{t^{\prime}}^{H}$, and $f_{t+t^{\prime}}^{H}$ exist.

A fundamental property of Hamiltonian flows is that each $f_{t}^{H}$ is a
canonical transformation (or symplectomorphism). This means that for every
$z=(x,p)$ the Jacobian matrix
\[
Df_{t}^{H}(z)=\frac{\partial z_{t}}{\partial z}=\frac{\partial(x_{t},p_{t}%
)}{\partial(x,p)}%
\]
is symplectic, i.e. $Df_{t}^{H}(z)\in\operatorname{Sp}(2n,\mathbb{R})$. (In
some texts the term \textquotedblleft canonical
transformation\textquotedblright\ refers more generally to mappings that
preserve the form of Hamilton's equations; the definition we use here is thus
more restrictive). The simplest proof of this property consists in showing
that the matrix $S_{t}=Df_{t}^{H}(z)$ satisfies the first order matrix
equation
\begin{equation}
\frac{d}{dt}S_{t}=JH^{\prime\prime}(z_{t},t)S_{t}\text{ } \label{vari}%
\end{equation}
where $H^{\prime\prime}=D^{2}H$ is the Hessian matrix of the Hamiltonian, i.e.
the matrix of its second derivatives in the variables $x$ and $p$ (see
\cite{Birk}, \S 2.3.2) for the derivation of the equation (\ref{vari}), which
is sometimes called the \textquotedblleft variational
equation\textquotedblright\ in the literature). To show that $S_{t}$ is
symplectic, we first set $A_{t}=(S_{t})^{T}JS_{t}.\,\ $It follows from this
equation that
\begin{align*}
\frac{dA_{t}}{dt}  &  =\frac{d(S_{t})^{T}}{dt}JS_{t}+(S_{t})^{T}%
J\,\frac{dS_{t}}{dt}\\
&  =(S_{t})^{T}H^{\prime\prime}(z_{t})S_{t}-(S_{t})^{T}H^{\prime\prime}%
(z_{t})S_{t}\\
&  =0
\end{align*}
hence $A_{t}$ is constant. Thus $A_{t}=A_{0}=(S_{0})^{T}JS_{0}=J$ which shows
that $S_{t}\in\operatorname{Sp}(2n,\mathbb{R})$.

Let now $s$ be an arbitrary symplectic matrix; by the chain rule, the Jacobian
matrix of $sf_{t}^{H}s^{-1}$ is $sS_{t}s^{-1}$ hence it is also a canonical
transformation; it is in fact the flow of a certain Hamiltonian. Let us now
construct this Hamiltonian:

\begin{proposition}
\label{propco}The family of canonical transformations $sf_{t}^{H}s^{-1}$ is
the Hamiltonian flow determined by the function $K(z,t)=H(s^{-1}z,t)$. That
is,%
\begin{equation}
sf_{t}^{H}s^{-1}=f_{t}^{K}\text{ , }K(z)=H(s^{-1}z). \label{flow1}%
\end{equation}

\end{proposition}

\begin{proof}
Writing Hamilton equations for $K$ as $\dot{z}=J\nabla_{z}[H(s^{-1}z,t)]$ we
get, using by the chain rule
\[
\nabla_{z}[H(s^{-1}z)]=(s^{T})^{-1}(\nabla_{z}H)(s^{-1}z)
\]
hence, since $sJs^{T}=J$ because $s$ is symplectic,%
\[
\dot{z}=J(s^{T})^{-1}(\nabla_{z}H)(s^{-1}z)=sJ(\nabla_{z}H)(s^{-1}z)].
\]
Thus $s^{-1}z(t)$ is the solution of Hamilton's equations for $H$ with initial
datum $s^{-1}z(0)$; it follows that $f_{t}^{H}(s^{-1}z)=s^{-1}f_{t}^{K}(z)$
which is equivalent to formula (\ref{flow1}).
\end{proof}

We will call the property just proved the \emph{symplectic covariance property
of Hamiltonian mechanics}. It will be generalized in the next subsection to
the nonlinear case, where we study products of Hamiltonian flows. Formula
(\ref{flow1}) remains true if we replace$\ s$ by an arbitrary symplectic
transformation $f$. The result is actually well-known in standard Hamiltonian
mechanics in the following form: set $(x^{\prime},p^{\prime})=f(x,p)$ and
$K=H\circ f$; if $f$ is a canonical transformation then we have the
equivalence
\begin{gather}
\dot{x}^{\prime}=\nabla_{p^{\prime}}K(x^{\prime},p^{\prime},t)\text{ \ and
}\dot{p}^{\prime}=-\nabla_{x^{\prime}}K(x^{\prime},p^{\prime},t)\text{\ }%
\nonumber\\
\Longleftrightarrow\label{equiha}\\
\dot{x}=\nabla_{p}H(x,p,t)\text{ \ and \ }\dot{p}=-\nabla_{x}%
H(x,p,t).\nonumber
\end{gather}
(see de Gosson \cite{Birk}, \S 2.3.2, for a proof). Let $X_{H}$ the Hamilton
vector field of $H$; by definition%
\[
X_{H}(z,t)=(\nabla_{p}H(z,t),-\nabla_{x}H(z,t)).
\]
Hamilton's equations are thus equivalent to $\dot{z}=X_{H}(z,t)$. Note that
$X_{H}(z,t)$ is not strictly speaking a vector field on $\mathbb{R}^{2n}$
because of the dependence in $t$. In terms of $X_{H}$ we can restate the
theorem above as the transformation law
\begin{equation}
X_{H\circ f}(z)=[Df(z)]^{-1}(X_{H}\circ f)(z) \label{ks}%
\end{equation}
for Hamilton vector fields.

\subsection{Operations on Hamiltonian Flows}

We assume as before that all Hamiltonians have flows defined for all values of
time $t$.

\begin{proposition}
\label{prop2}Let $(f_{t}^{H})$ and $(f_{t}^{K})$ be Hamiltonian flows. Then:%
\begin{align}
f_{t}^{H}f_{t}^{K}  &  =f_{t}^{H\#K}\text{ \ \ with \ \ }%
H\#K(z,t)=H(z,t)+K((f_{t}^{H})^{-1}(z),t).\label{ch1}\\
(f_{t}^{H})^{-1}  &  =f_{t}^{K}\text{ \ \ with \ \ }K(z,t)=-H(f_{t}^{H}(z),t).
\label{ch2}%
\end{align}

\end{proposition}

\begin{proof}
Let us first prove (\ref{ch1}). Using successively the product and chain
rules, we have%
\begin{align*}
\frac{d}{dt}(f_{t}^{H}f_{t}^{K})  &  =\left(  \frac{d}{dt}f_{t}^{H}\right)
f_{t}^{K}+(Df_{t}^{H})f_{t}^{K}\frac{d}{dt}f_{t}^{K}\\
&  =X_{H}(f_{t}^{H}f_{t}^{K})+(Df_{t}^{H})f_{t}^{K}\circ X_{K}(f_{t}^{K})
\end{align*}
and it thus suffices to show that
\begin{equation}
(Df_{t}^{H})f_{t}^{K}\circ X_{K}(f_{t}^{K})=X_{K\circ(f_{t}^{H})^{-1}}%
(f_{t}^{K})\text{.} \label{fksuf}%
\end{equation}
Writing%
\[
(Df_{t}^{H})f_{t}^{K}\circ X_{K}(f_{t}^{K})=(Df_{t}^{H})((f_{t}^{H})^{-1}%
f_{t}^{H}f_{t}^{K})\circ X_{K}((f_{t}^{H})^{-1}f_{t}^{H}f_{t}^{K})
\]
the equality (\ref{fksuf}) follows from the transformation formula (\ref{ks})
for Hamilton vector fields. Formula (\ref{ch2}) is now an easy consequence of
(\ref{ch1}), noting that if $K$ is given by the second formula (\ref{ch2})
then $(f_{t}^{H}f_{t}^{K})$ is the flow determined by the Hamiltonian
\[
K(z,t)=H(z,t)+K((f_{t}^{H})^{-1}(z),t)=0;
\]
$f_{t}^{H}f_{t}^{K}$ is thus the identity, so that $(f_{t}^{H})^{-1}=f_{t}%
^{K}$ as claimed.
\end{proof}

The formulae above show why we cannot avoid considering time-dependent
Hamiltonians: even if $H$ and $K$ do not depend explicitly on time, the
product and inverse of their flows is generated by time-dependent
Hamiltonians! This remark will be important below when we define the group of
Hamiltonian canonical transformations.

\subsection{The group of Hamiltonian canonical
transformations\label{subsecham}}

From now on we assume that the solutions for the Hamilton equations with
arbitrary initial data at time $t=0$ are uniquely determined by these data and
exist for all times. This may seem to be a strong technical restriction, but
it can actually easily be implemented (see e.g. Polterovich \cite{Polter}). If
$H$ is such that $f_{t}^{H}$ is not defined for all values of $t$, we just
replace $H$ by the function $H\Theta$ where $\Theta=\Theta(x,p)$ is a
compactly supported infinitely differentiable function equal to one on some
arbitrarily chosen subset $\Omega$ of phase space.

Now proposition \ref{prop2} allows us to define the group $\operatorname{Ham}%
(2n,\mathbb{R})$ of Hamiltonian canonical transformations. To do this, let us
introduce the following terminology. We will say that a canonical
transformation $f$ of $\mathbb{R}^{2n}$ is \emph{Hamiltonian} if it is the
time-one flow of some Hamiltonian function, i.e. if there exists a
(time-dependent) Hamiltonian $H$ such that we have $f=f_{t=1}^{H}$. Note the
set of all Hamiltonian canonical transformation of $\mathbb{R}^{2n}$ is
denoted by $\operatorname{Ham}(2n,\mathbb{R})$. The choice $t=1$ is arbitrary,
and can be replaced by any other time $t_{0}$, as a consequence of the
following simple observation: define the Hamiltonian function $K(z,t)=t_{0}%
H(z,t_{0}t)$; the flow it determines is given by $f_{t/t_{0}}^{K}=f_{t}^{H}$,
hence $f_{t_{0}}^{H}=f_{1}^{K}$ is also a Hamiltonian canonical transformation.

It follows from Proposition \ref{prop2} that $\operatorname{Ham}%
(2n,\mathbb{R})$ is indeed a group for the product operation: first, the
product is clearly associative (because composition of mappings is). Assume
that $H$ is a constant; then $f_{t}^{H}$ is the identity $I$ on $\mathbb{R}%
^{2n}$ and hence $I$ is in $\operatorname{Ham}(2n,\mathbb{R})$. Suppose now
$f$ and $g$ are two Hamiltonian canonical transformations: $f=f_{t=1}^{H}$ and
$g=f_{t=1}^{K}$. In view of formula (\ref{ch1}) we have $fg=f_{t=1}^{H\#K}$
hence $fg$ is also a Hamiltonian canonical transformation. That the inverse
$f^{-1}$ also is in $\operatorname{Ham}(2n,\mathbb{R})$ follows similarly from
formula (\ref{ch2}).

\subsection{The key role of Banyaga's Theorem.}

The group $\operatorname{Ham}(2n,\mathbb{R})$ contains the symplectic group
$\operatorname{Sp}(2n,\mathbb{R})$ as a subgroup. Equivalently and more
important for us here: \emph{every symplectic matrix is the time-one flow of
some Hamiltonian function}. This is obvious if $s$ is in the range of the
exponential $\exp:\mathfrak{sp}(2n,\mathbb{R})\longrightarrow\operatorname{Sp}%
(2n,\mathbb{R})$, writing $s=\exp(X)$: the formula $s_{t}=\exp(tX)$ then
defines a one-parameter group of symplectic matrices, which is the propagator
for the Hamilton equations for $H(z)=-\frac{1}{2}JXz\cdot z$. The general case
easily follows by induction since every symplectic matrix can be written as a
product $\exp(X_{1})\cdot\cdot\cdot\exp(X_{k})$ where $X_{1},...,X_{k}$ are in
the Lie algebra $\mathfrak{sp}(2n,\mathbb{R})$. There is actually a more
constructive way to prove this result as a particular case of the following
far-reaching property of $\operatorname{Ham}(2n,\mathbb{R})$, proven in full
generality by Banyaga \cite{Banyaga} in the context of symplectic manifolds:

\begin{theorem}
[Banyaga]Let $(f_{t})$ be a family of Hamiltonian canonical transformations
depending smoothly on the parameter $t$, and such that $f_{0}=I$ (the
identity). Then $f_{t}=f_{t}^{H}$ for some Hamiltonian function $H(z,t)$ given
by the formula%
\begin{equation}
H(z,t)=H(z,0)-\int_{0}^{1}\sigma(X(uz,t),z)du \label{hzt}%
\end{equation}
where $X=\left(  \frac{d}{dt}f_{t}\right)  f_{t}^{-1}$ and $\sigma$ is the
symplectic form $\sigma(z,z^{\prime})=(z^{\prime})^{T}Jz=Jz\cdot z^{\prime}$
on the phase space $\mathbb{R}^{2n}$.
\end{theorem}

Let us sketch the proof (for a complete proof see Polterovich \cite{Polter} or
Banyaga's original paper \cite{Banyaga}). One begins by noting that if $X_{H}$
is a Hamiltonian vector field, one can reconstruct $H$ by the following
method: first write%
\begin{align*}
H(z,t)  &  =H(z,0)+\int_{0}^{1}\frac{d}{du}X_{H}(uz,t)du\\
&  =H(z,0)+\int_{0}^{1}\left[  \nabla_{z}H(uz,t)\cdot z\right]  du
\end{align*}
(the second equality in view of the chain rule). Next observe that since
$\nabla_{z}H(uz,t)=-J^{2}\nabla_{z}H(uz,t)=-JX_{H}(uz,t)$, we have%
\[
H(z,t)=H(z,0)-\int_{0}^{1}\sigma(X_{H}(uz,t),z)du
\]
where $\sigma$ is the standard symplectic form. One then proves the
proposition by setting $X=\left(  \frac{d}{dt}f_{t}\right)  f_{t}^{-1}$ and
showing that the flow determined by the Hamiltonian functions is precisely
$(f_{t})$.

This result is remarkable, and even in some sense surprising, because it shows
that the datum of a family of time-one Hamiltonian canonical transformations,
each coming from a different Hamiltonian function, is itself the flow of some Hamiltonian.

This has several non-trivial consequences. First, it shows that the group
$\operatorname{Ham}(2n,\mathbb{R})$ is arcwise connected; secondly it
immediately implies that \newline$\operatorname{Sp}(2n,\mathbb{R}%
)\subset\operatorname{Ham}(2n,\mathbb{R})$: let $s\in\operatorname{Sp}%
(2n,\mathbb{R})$ and choose a path $(s_{t})$ joining the identity to $s$ in
$\operatorname{Sp}(2n,\mathbb{R})$. Then $X=\left(  \frac{d}{dt}s_{t}\right)
s_{t}^{-1}$ belongs to the symplectic Lie algebra $\mathfrak{sp}%
(2n,\mathbb{R})$; in particular it is a matrix so that formula (\ref{hzt})
defines a quadratic Hamiltonian whose associated flow is precisely $(s_{t})$.

\subsection{Reduction to the time-independent case: the extended phase space}

It turns out that the study of time-dependent Hamiltonian function can always
be reduced to that of Hamiltonians which do not explicitly contain time. As we
will see, the price we have to pay for this is that we have to work in a phase
space with dimension $n+2$ instead of $n$. Thus, in a sense, there are no
time-dependent Hamiltonians!

The trick is the following, which seems to go back to \cite{synge,lan} (see
\cite{struck} for an analysis in depth). We define a new Hamiltonian function
$\widetilde{H}$ by the formula%
\begin{equation}
\widetilde{H}(x,p,t,E)=H(x,p,t)-E \label{htilda}%
\end{equation}
where $E$ is a new variable, viewed as conjugate to the time $t$ (the latter
now has the status of a \textquotedblleft position variable\textquotedblright;
we could as well write the definition of $\widetilde{H}$ in the form
\[
\widetilde{H}(x,p,x_{n+1},p_{n+1})=H(x,p,x_{n+1})-p_{n+1}%
\]
but we will however stick to the notation (\ref{htilda}) if only for economy.
The function $\widetilde{H}$ is defined on the extended phase space
$\mathbb{R}^{2n+2}\equiv\mathbb{R}_{x,p}^{2n}\times\mathbb{R}_{E}%
\times\mathbb{R}_{t}$ and the associated Hamilton equations are, expressed in
terms of the original Hamiltonian $H$:%
\begin{align*}
\frac{dx}{dt^{\prime}}  &  =\nabla_{p}H\text{ \ , \ }\frac{dp}{dt^{\prime}%
}=-\nabla_{x}H\text{ }\\
\frac{dE}{dt^{\prime}}  &  =\frac{\partial H}{\partial t}\text{ \ , \ }%
\frac{\partial t}{\partial t^{\prime}}=1
\end{align*}
where the parameter $t^{\prime}$ plays the role of a new \textquotedblleft
time\textquotedblright; since $\widetilde{H}$ does not contain explicitly that
parameter it is indeed a \textquotedblleft time-independent\textquotedblright%
\ Hamiltonian on extended phase space. Notice that in view of the fourth
equation above we may choose $t^{\prime}=t$ so that the two first equations
are just the Hamiltonian equations for $H$; as a bonus the third equation is
just the familiar law for the variation of energy of a time-dependent
Hamiltonian system:%
\[
\frac{dE}{dt}=\frac{d}{dt}H(x,p,t)=\frac{\partial}{\partial t}H(x,p,t)
\]
(the second equality because of the chain rule and using the fact that $x$ and
$p$ satisfy Hamilton's equations).

We can now define the \textquotedblleft extended Hamiltonian
flow\textquotedblright\ $(\widetilde{f}_{t}^{H})$ of $H$ by the formula
$\widetilde{f}_{t}^{H}=f_{t}^{\widetilde{H}}$. Notice that since
$(f_{t}^{\widetilde{H}})$ is the flow determined by a time-independent
Hamiltonian $(\widetilde{f}_{t}^{H})$, it enjoys the one-parameter group
property $\widetilde{f}_{t}^{H}\widetilde{f}_{t^{\prime}}^{H}=\widetilde{f}%
_{t+t^{\prime}}^{H}$ and $\widetilde{f}_{0}^{H}=I$ \ (the identity operator on
the extended phase space $\mathbb{R}^{2n+2}$).

Denote now by $(f_{t,t^{\prime}}^{H})$ the two-parameter family of canonical
transformations of $\mathbb{R}^{2n}$ defined as follows: for fixed $t^{\prime
}$ the function $z=f_{t,t^{\prime}}^{H}(z^{\prime})$ is the solution of
Hamilton's equations for $H$ taking the value $z^{\prime}$ at time $t^{\prime
}$. Clearly $f_{t,t}^{H}$ is the identity operator on $\mathbb{R}^{2n}$ and
$f_{t,t^{\prime}}^{H}f_{t^{\prime},t^{\prime\prime}}^{H}=f_{t,t^{\prime\prime
}}^{H}$, $(f_{t,t^{\prime}}^{H})^{-1}=f_{t^{\prime},t}^{H}$. The two-parameter
family $(f_{t,t^{\prime}}^{H})$ is sometimes called the \textquotedblleft
time-dependent flow\textquotedblright; it is related to the extended flow
defined above by the simple (and useful) formula%
\begin{equation}
\widetilde{f}_{t}^{H}(z^{\prime},t^{\prime},E^{\prime})=(f_{t,t^{\prime}}%
^{H}(z^{\prime}),t+t^{\prime},E_{t,t^{\prime}}) \label{ftze}%
\end{equation}
with
\begin{equation}
E_{t,t^{\prime}}=E^{\prime}+H(f_{t,t^{\prime}}^{H}(z^{\prime}),t)-H(z^{\prime
},t^{\prime}). \label{energy}%
\end{equation}
We will see below that the procedure above extends, suitably modified, to the
case of the Schr\"{o}dinger equation.

\section{The Weyl Formalism and Schr\"{o}dinger's Equation}

In classical physics, observables are taken to be functions on phase space,
limited by whatever conditions of regularity that are required by the physics
of the problem. A class of observables having been chosen, a state is a
continuous functional on this class (the continuity being imposed for some
specified topology). We will cast a wide net and accept as observable not only
functions, but any tempered distribution on the configuration space
$\mathbb{R}^{d}$; the space of such distributions is traditionally denoted by
$\mathcal{S}^{\prime}(\mathbb{R}^{n})$. It is the dual of the Schwartz space
$\mathcal{S}(\mathbb{R}^{n})$ of all complex functions defined on
$\mathbb{R}^{n}$ which decrease, together with their derivatives, faster than
the inverse of any polynomial in the variables $x_{1},...,x_{n}$.

An observable will usually be denoted by a small Latin letter, such that
$a,b,$...; tradition provides us with the exception confirming the rule: when
an observable is viewed as a Hamiltonian function it will always be denoted by
$H,K$, etc.

\subsection{Pseudodifferential operators}

We begin by briefly discussing some well-known aspects from the theory of
pseudodifferential operators (see for instance Shubin \cite{Shubin}). Let $A$
be a linear operator $\mathcal{S(}\mathbb{R}^{n})\longrightarrow
\mathcal{S}^{\prime}(\mathbb{R}^{n})$. If we assume that $A$ is continuous
from $\mathcal{S}(\mathbb{R}^{n})$ to $\mathcal{S}^{\prime}(\mathbb{R}^{n})$
(when $\mathcal{S}^{\prime}(\mathbb{R}^{n})$ is equipped with the weak-*
topology, that is the topology of pointwise convergence) then a deep result
from functional analysis, Schwartz's kernel theorem, tells us that there
exists a distribution $K_{A}$ in $\mathcal{S}^{\prime}(\mathbb{R}^{n}%
\times\mathbb{R}^{n})$ such that for all $\psi,\phi$ in $\mathcal{S}%
(\mathbb{R}^{n})$, we have $\langle A\psi,\phi\rangle=$ $\langle\langle
K_{A},\psi\otimes\phi^{\ast}\rangle\rangle$ where $\langle\cdot,\cdot\rangle$
and $\langle\langle\cdot,\cdot\rangle\rangle$ are the distributional brackets
for $\mathcal{S}^{\prime}(\mathbb{R}^{n})$ and $\mathcal{S}^{\prime
}(\mathbb{R}^{n}\times\mathbb{R}^{n})$, respectively (see Gr\"{o}chenig
\cite{Gro}, \S 14.3, for an elegant proof of the kernel theorem in the broader
context of Feichtinger's modulation spaces). We will informally write%
\begin{equation}
A\psi(x)=\int_{\mathbb{R}^{n}}K_{A}(x,y)\psi(y)dy \label{akernel}%
\end{equation}
where the integral should be interpreted in the distributional sense.

Let now $\tau$ be an arbitrary fixed real number and define a distribution
$a_{\tau}(x,p)$ by the formula%
\begin{equation}
a_{\tau}(x,p)=\int_{\mathbb{R}^{n}}e^{-\frac{i}{\hbar}p\cdot y}K_{A}(x+\tau
y,x-(1-\tau)y)dy; \label{atau}%
\end{equation}
where we have introduced a scaling factor which for convenience, we have
called $\hbar$. This ensures that the exponential is dimensionless. The
magnitude of this factor is left open in this paper since we are concentrating
only on mathematical relationships.

Using the Fourier inversion formula, the kernel is then expressed in terms of
$a$ by%
\begin{equation}
K_{A}(x,y)=\left(  \tfrac{1}{2\pi\hbar}\right)  ^{n}\int_{\mathbb{R}^{n}%
}e^{\frac{i}{\hbar}p\cdot(x-y)}a_{\tau}((1-\tau)x+\tau y,p)dp. \label{ka}%
\end{equation}
It follows that formula (\ref{akernel}) can be rewritten in pseudodifferential
form as
\begin{equation}
A\psi(x)=\left(  \tfrac{1}{2\pi\hbar}\right)  ^{n}\iint\nolimits_{\mathbb{R}%
^{n}\times\mathbb{R}^{n}}e^{\frac{i}{\hbar}p\cdot(x-y)}a_{\tau}((1-\tau)x+\tau
y,p)\psi(y)dydp \label{taupdo}%
\end{equation}
Thus we have shown how to associate an operator $A$ with an observable
$a_{\tau}$.

Now we can use formula (\ref{taupdo}) to associate to given observable $a$
with an operator $A_{\tau}$ for every value of the constant $\tau$: by
definition%
\begin{equation}
A_{\tau}\psi(x)=\left(  \tfrac{1}{2\pi\hbar}\right)  ^{n}\iint%
\nolimits_{\mathbb{R}^{n}\times\mathbb{R}^{n}}e^{\frac{i}{\hbar}p\cdot
(x-y)}a((1-\tau)x+\tau y,p)\psi(y)dydp \label{Atau}%
\end{equation}
the \textquotedblleft double integral\textquotedblright\ being again
interpreted in some reasonable sense (for instance as a repeated
distributional bracket; it is actually no restriction in our discussion to
assume that the observable is in $\mathcal{S}(\mathbb{R}^{2n})$ in which case
the right hand side of (\ref{Atau}) strictly makes sense since the integral is
absolutely convergent).

Let us dignify the discussion above in compact form as a Theorem:

\begin{theorem}
\label{every}Every continuous linear operator $A:\mathcal{S}(\mathbb{R}%
^{n})\longrightarrow\mathcal{S}^{\prime}(\mathbb{R}^{n})$ can be written in
pseudodifferential form (\ref{taupdo}), and this for every real value of the
parameter $\tau$.
\end{theorem}

The (admittedly somewhat technical) discussion above shows that for each value
of $\tau$ we have a bijective (= one-to-one and onto) correspondence
$a\longleftrightarrow A_{\tau}$ between observables and continuous operators
$\mathcal{S}(\mathbb{R}^{n})\longrightarrow\mathcal{S}^{\prime}(\mathbb{R}%
^{n})$ given by (\ref{Atau}). We emphasize again that each choice of $\tau$ is
a priori equally good; for instance if we take $\tau=0$ then we get%
\begin{align*}
A_{1}\psi(x)  &  =\left(  \tfrac{1}{2\pi\hbar}\right)  ^{n}\iint%
\nolimits_{\mathbb{R}^{n}\times\mathbb{R}^{n}}e^{\frac{i}{\hbar}p\cdot
(x-y)}a(x,p)\psi(y)dydp\\
&  =\left(  \tfrac{1}{2\pi\hbar}\right)  ^{n}\int_{\mathbb{R}^{n}}e^{\frac
{i}{\hbar}p\cdot x}a(x,p)F\psi(p)dy
\end{align*}
($F\psi$ the $\hbar$-Fourier transform) which is the standard
pseudodifferential representation widely used in partial differential
equations. There is however a privileged choice in quantum mechanics: it
consists in taking $\tau=\frac{1}{2}$. The corresponding operator $A_{1/2}$
will be denoted $\widehat{A}$ and called the Weyl operator corresponding to
the observable (or symbol) $a$. Explicitly%
\[
\widehat{A}\psi(x)=\left(  \tfrac{1}{2\pi\hbar}\right)  ^{n}\iint%
\nolimits_{\mathbb{R}^{n}\times\mathbb{R}^{n}}e^{\frac{i}{\hbar}p\cdot
(x-y)}a(\tfrac{1}{2}(x+y),p)\psi(y)dydp
\]
and a few elementary manipulations show that this equivalent to the following
expression%
\[
\widehat{A}\psi(x)=\left(  \tfrac{1}{2\pi\hbar}\right)  ^{n}\int%
_{\mathbb{R}^{n}}a_{\sigma}(z^{\prime})\widehat{T}(z^{\prime})\psi
(x)dz^{\prime}%
\]
which is more familiar to physicist; here
\[
a_{\sigma}(x^{\prime},p^{\prime})=\left(  \tfrac{1}{2\pi\hbar}\right)
^{n}\int_{\mathbb{R}^{2n}}e^{-\frac{i}{\hbar}\sigma(z^{\prime},z^{\prime
\prime})}a(z^{\prime\prime})dz^{\prime\prime}%
\]
(the symplectic Fourier transform of $a$) and $\widehat{T}(z^{\prime})$ is the
Heisenberg--Weyl operator:%
\[
\widehat{T}(z^{\prime})\psi(x)=e^{\frac{i}{\hbar}(p^{\prime}\cdot x-\frac
{1}{2}p^{\prime}\cdot x^{\prime})}\psi(x-x^{\prime});
\]
see Littlejohn \cite{Littlejohn} for an analysis of these operators of which
we have given a detailed study in \cite{Birk}.

The Weyl correspondence associates in a unique way to every observable
$a\in\mathcal{S}^{\prime}(\mathbb{R}^{n})$ a continuous linear operator
$\widehat{A}$ whose domain always contains $\mathcal{S}(\mathbb{R}^{n})$ and
whose range is contained in $\mathcal{S}^{\prime}(\mathbb{R}^{n})$. This
correspondence actually is not only one-to-one, but also onto: every
continuous linear operator $\widehat{A}:\mathcal{S}(\mathbb{R}^{n}%
)\longrightarrow\mathcal{S}^{\prime}(\mathbb{R}^{n})$ is associated to an
observable $a$ via the Weyl correspondence. We will symbolically write the
Weyl correspondence between an observable $a$ and the corresponding operator
$\hat A$ is $a\overset{\text{Weyl}}{\leftrightarrow}\widehat{A}$ or
$\widehat{A}\overset{\text{Weyl}}{\leftrightarrow}a$. (In the mathematical
literature $a$ is called the \textit{Weyl symbol }of $\widehat{A}$\;).

The Weyl correspondence has the following very agreeable feature, which
distinguishes it from other $\tau$-correspondences, and which already
motivated Weyl's interest: the transition to the adjoint operator is
particularly simple. If $a\overset{\text{Weyl}}{\leftrightarrow}\widehat{A}$
then $\overline{a}\overset{\text{Weyl}}{\leftrightarrow}\widehat{A}^{\ast}$.
In particular, if $a$ is a real observable then $\widehat{A}$ is formally
self-adjoint. Thus the Weyl correspondence associates to real observables
formally self-adjoint operators. This is one of the reasons why the
\textquotedblleft Weyl ordering\textquotedblright\ was adopted at an early
stage (but more or less explicitly) in quantum mechanics. The second reason is
that Weyl operators are characterized by a symplectic covariance property, an
important characterisation for the discussions in this paper. We will now
discuss the details of this symplectic covariance.

\subsection{Symplectic covariance of Weyl calculus}

The fundamental result, which is at the very heart of our derivation of
Schr\"{o}dinger's equation in next subsection is the following:

\begin{theorem}
\label{TheoCo}Let $\mathcal{L}(\mathcal{S}(\mathbb{R}^{n}),\mathcal{S}%
^{\prime}(\mathbb{R}^{n}))$ be the space of all continuous linear mappings
$\mathcal{S}(\mathbb{R}^{n})\longrightarrow\mathcal{S}^{\prime}(\mathbb{R}%
^{n})$. Let \
\[
\mathcal{M}:\mathcal{S}^{\prime}(\mathbb{R}^{2n})\longrightarrow
\mathcal{L}(\mathcal{S}(\mathbb{R}^{n}),\mathcal{S}^{\prime}(\mathbb{R}^{n}))
\]
be a continuous mapping associating to each observable $a$ a continuous
operator $\mathcal{M}(a)$. If $\mathcal{M}$ has the two following properties:

(i) If $a=a(x)$ then $\mathcal{M}(a)$ is the operator of multiplication by
$a(x)$;

(ii) For\textit{ every }$s$\textit{ in }$\operatorname*{Sp}(2n,\mathbb{R}%
)$\textit{ we have} $\mathcal{M}(a\circ s^{-1})=S\mathcal{M}(a)S^{-1}$
\textit{where }$S$\textit{ in} $\operatorname*{Mp}(2n,\mathbb{R})$ \textit{is
such that }$\pi^{\operatorname{Mp}}(S)=s$ then $\mathcal{M}$ is the Weyl
correspondence, that is $a\overset{\text{Weyl}}{\longleftrightarrow
}\mathcal{M}(a)$.
\end{theorem}

Note that if $a\overset{\text{Weyl}}{\leftrightarrow}\widehat{A}$ then we have
$a\circ s^{-1}\overset{\text{Weyl}}{\longleftrightarrow}S\widehat{A}S^{-1}$:
this is the well-known property of symplectic covariance of Weyl calculus; we
have given a detailed proof of this property in \cite{Birk}, \S 7.1.3. There
is also a proof in Littlejohn \cite{Littlejohn}.

The real impact of the theorem above is that it says that, conversely, among
all possible pseudodifferential correspondences $a\longleftrightarrow A_{\tau
}$ between observables and continuous operators (\ref{Atau}), the Weyl
correspondence is the \emph{only} one enjoying the property of symplectic
covariance. The proof of this result is alluded to in Stein's book
\cite{Stein} (pp. 578--579) and proven in detail in the last Chapter of Wong's
book \cite{Wong}. Wong's proof actually relies on the following additional
assumption (which is also found in Stein's statement): the mapping
$\mathcal{M}$ should associate to every essentially bounded observable $a$
which depends only on $x$, the operation of multiplication by $a$: if $a=a(x)$
then $Q(a)\psi=a\psi$. However this property is automatically satisfied by the
operators $A_{\tau}$: we have%
\begin{align*}
A_{\tau}\psi(x)  &  =\left(  \tfrac{1}{2\pi\hbar}\right)  ^{n}\iint%
\nolimits_{\mathbb{R}^{n}\times\mathbb{R}^{n}}e^{\frac{i}{\hbar}p\cdot
(x-y)}a((1-\tau)x+\tau y)\psi(y)dydp\\
&  =\left(  \tfrac{1}{2\pi\hbar}\right)  ^{n}\int_{\mathbb{R}^{n}}\left[
\int_{\mathbb{R}^{n}}e^{\frac{i}{\hbar}p\cdot(x-y)}dp\right]  a((1-\tau)x+\tau
y)\psi(y)dy\\
&  =\int_{\mathbb{R}^{n}}\delta(x-y)a((1-\tau)x+\tau y)\psi(y)dy\\
&  =a(x)\psi(x).
\end{align*}

\section{Derivation of Schr\"{o}dinger's equation}

\subsection{Statement of the main result}

Now we come to the main result in the paper, namely, to show that the
Schr\"{o}dinger equation can be derived from Hamilton's equations of motion by
lifting the flows $F_{t}^{H}$ of Ham($2n, \mathbb{R}$) onto a unitary
representation of a covering structure.

In order to motivate our approach let us now briefly return to the property of
the metaplectic representation of $\operatorname{Sp}(2n,\mathbb{R})$ which
shows that to every family $(s_{t})$ of symplectic matrices depending smoothly
on $t$ and such that $s_{0}=I$, we can associate, in a unique way, a family
$(S_{t})$ of unitary operators on $L^{2}(\mathbb{R}^{n})$ belonging to the
metaplectic group $\operatorname{Mp}(2n,\mathbb{R})$ such that $S_{0}=1$ and
$S_{t}S_{t^{\prime}}=S_{t+t^{\prime}}$.

We have seen from Banyaga's theorem, that for every flow $(s_{t})$ there
exists some Hamiltonian $H$, so that we can write $(s_{t})=(f_{t}^{H})$. Now
if we use the correspondence $H\overset{\text{Weyl}}{\leftrightarrow
}\widehat{H}$ we have $(S_{t})=(F_{t}^{H})$ where
\[
S_{t}=e^{-i\widehat{H}t/\hbar}%
\]
which is the solution of the Schr\"{o}dinger equation
\[
i\hbar\frac{d}{dt}S_{t}=\widehat{H}S_{t}%
\]
Here we have again introduced a scaling parameter having the dimensions of
action which, for convenience, we have denoted by $\hbar$. Note that
generically $\widehat{H}$ is not a bounded operator on $L^{2}(\mathbb{R}^{n})$
so that the exponential has to be defined using some functional calculus (see
Reed and Simon \cite{RS} \S VIII.3 for a discussion of these technicalities).

We now ask whether this property has an analogue for paths in the group
$\operatorname{Ham}(2n,\mathbb{R})$ of Hamiltonian canonical transformations.
That is can we find a $(F_{t}^{H})$ corresponding to $(f_{t}^{H})$ for every
physically relevant Hamiltonian. To sharpen up the discussion, let us
introduce the following notation:

\begin{itemize}
\item Denote by $\mathcal{P}\operatorname{Ham}(2n,\mathbb{R})$ the set of all
one-parameter families $(f_{t})$ in $\operatorname{Ham}(2n,\mathbb{R})$
depending smoothly on $t$ and passing through the identity at time $t=0$. As
we have remarked above, such a family of canonical transformations is always
the flow $(f_{t}^{H})$ of some (usually time-dependent) Hamiltonian $H$
(Banyaga's theorem).

\item Denote by $\mathcal{P}U(L^{2}(\mathbb{R}^{n}))$ the set of all strongly
continuous one-parameter families $(F_{t})$ of unitary operators on
$L^{2}(\mathbb{R}^{n})$ depending smoothly on $t$ and such that $F_{0}$ is the
identity operator, and having the following property: the domain of the
infinitesimal generator $\widehat{H}$ of $(F_{t})$ contains the Schwartz space
$\mathcal{S}(\mathbb{R}^{n})$.
\end{itemize}

Recall that strong continuity for a one-parameter group $(F_{t})$ means that
we have
\begin{equation}
\lim_{t\rightarrow t_{0}}F_{t}\psi=F_{t_{0}}^{H}\psi\label{strong1}%
\end{equation}
and the infinitesimal generator of $(F_{t})$ is the the operator defined by
\begin{equation}
A=i\hbar\left.  \frac{d}{dt}F_{t}\psi\right\vert _{t=0}=i\hbar\lim_{\Delta
t\rightarrow0}\frac{F_{\Delta t}\psi-\psi}{\Delta t} \label{Hhat}%
\end{equation}
for every $\psi\in L^{2}(\mathbb{R}^{n})$ and every real number $t_{0}$;
formally $F_{t}=e^{-i\hbar A/t}$. Formally we have Stone's theorem
\cite{Stone}:

\begin{theorem}
[Stone]For every strongly continuous one-parameter group $(F_{t})$ of unitary
operators on a Hilbert space $\mathcal{H}$ there exists a self-adjoint
operator $A$ on $L^{2}(\mathbb{R}^{n})$ such that $F_{t}=e^{itA/\hbar}$; in
particular $A$ is closed and densely defined in $\mathcal{H}$. Conversely, if
$A$ is a self-adjoint operator on $\mathcal{H}$ then there exists a unique
one-parameter unitary group $(F_{t})$ whose infinitesimal generator is $A$,
that is $F_{t}=e^{itA/\hbar}$.
\end{theorem}

For self-contained proofs we refer Reed and Simon \cite{RS}, \S VIII.4, or to
Abraham et al. \cite{AMR}, Supplement 7.4B, pp.529--535.

We will now use Stone's theorem to prove the hard part of our main result:

\begin{theorem}
\label{MC}There exists a bijective correspondence
\[
\mathcal{C}:\mathcal{P}U(L^{2}(\mathbb{R}^{n}))\leftrightarrow\mathcal{P}%
\operatorname{Ham}(2n,\mathbb{R})
\]
whose restriction to families $(s_{t})$ of symplectic matrices reduces to the
metaplectic representation, and which has the following symplectic covariance
property: for every $(f_{t})$ in $\mathcal{P}\operatorname{Ham}(2n,\mathbb{R}%
)$ and for every $s$ in $\operatorname{Sp}(2n,\mathbb{R})$ we have%
\begin{equation}
\mathcal{C}(sf_{t}s^{-1})=(SF_{t}S^{-1}) \label{symco}%
\end{equation}
where $S$ is any of the two operators in $\operatorname{Mp}(2n,\mathbb{R})$
such that $\pi^{\operatorname{Mp}}(S)=s$. This correspondence $\mathcal{C}$ is
bijective and we have
\begin{equation}
i\hbar\frac{d}{dt}F_{t}=\widehat{H}F_{t}\text{.} \label{sch1}%
\end{equation}
where $\widehat{H}\overset{\text{Weyl}}{\longleftrightarrow}H$, the
Hamiltonian function $H$ being determined by $(f_{t})$.
\end{theorem}

It is perhaps worth observing that it is always preferable to take the family
$(F_{t})$ as the fundamental object, rather than $\widehat{H}$ (and hence
Schr\"{o}dinger's equation). This remark has already been made by Weyl who
noticed that $(F_{t})$ is everywhere defined and consists of bounded
operators, while $\widehat{H}$ is generically unbounded and only densely
defined (see the discussion in Mackey \cite{Mackey} for a discussion of
related questions).

\subsection{Proof: the time-independent case}

Let us begin with the case where $(f_{t})$ is the Hamiltonian flow determined
by a time-independent Hamiltonian function $H=H(x,p)$, in which case
$(f_{t})=(f_{t}^{H})$ is a one-parameter group, that is $f_{t}^{H}%
f_{t^{\prime}}^{H}=f_{t+t^{\prime}}^{H}$. We thus want to associate to
$(f_{t}^{H})$ a strongly continuous one-parameter group $(F_{t})=(F_{t}^{H})$
of unitary operators on $L^{2}(\mathbb{R}^{n})$ satisfying some additional
conditions. We proceed as follows: let $\widehat{H}$ be the operator
associated to $H$ by the Weyl correspondence: $\widehat{H}\overset{\text{Weyl}%
}{\longleftrightarrow}H$ and define $\mathcal{C}(f_{t}^{H})=(F_{t})$ by
$F_{t}=e^{-it\widehat{H}/\hbar}$. The Weyl operator $\widehat{H}$ is
self-adjoint and its domain obviously contains $\mathcal{S}(\mathbb{R}^{n})$.
Let us show that the covariance property (\ref{symco}) holds. We have
\[
\mathcal{C}(sf_{t}^{H}s^{-1})=\mathcal{C}(f_{t}^{H\circ s^{-1}})
\]
in view of formula (\ref{flow1}) in Proposition \ref{propco}, that is, by
definition of $\mathcal{C}$,
\[
\mathcal{C}(sf_{t}^{H}s^{-1})=(e^{-it\widehat{H\circ s^{-1}}/\hbar}).
\]
In view of the symplectic covariance property $a\circ s^{-1}%
\overset{\text{Weyl}}{\longleftrightarrow}S\widehat{A}S^{-1}$ of Weyl
operators we have $\widehat{H\circ s^{-1}}=S\widehat{H}S^{-1}$, and hence%
\[
\mathcal{C}(sf_{t}^{H}s^{-1})=(e^{-itS\widehat{H}S^{-1}/\hbar}%
)=(Se^{-it\widehat{H}/\hbar}S^{-1})
\]
which is property (\ref{symco}).

Let conversely $(F_{t})$ be in $\mathcal{P}U(L^{2}(\mathbb{R}^{n}))$; we must
show that we can find a unique $(f_{t})$ in $\mathcal{P}\operatorname{Ham}%
(2n,\mathbb{R})$ such that $\mathcal{C}(f_{t})=(F_{t})$. By Stone's theorem
and our definition of $\mathcal{P}U(L^{2}(\mathbb{R}^{n}))$ there exists a
unique self-adjoint operator $A$, densely defined, and whose domain contains
$\mathcal{S}(\mathbb{R}^{n})$. Thus $A$ is continuous on $\mathcal{S}%
(\mathbb{R}^{n})$ and in view of Theorem \ref{every} for each value of the
parameter $\tau$ there exists an observable $a$ such that
$A\longleftrightarrow a_{\tau}$. Choose $\tau=\frac{1}{2}$; then
$A=\widehat{H}\overset{\text{Weyl}}{\longleftrightarrow}H$ for some function
$H=H(x,p)$ and we have $\mathcal{C}(f_{t}^{H})=(F_{t})$.

There remains to show that the correspondence $\mathcal{C}$ restricts to the
metaplectic representation for semigroups $(f_{t})=(s_{t})$ in
$\operatorname{Sp}(2n,\mathbb{R)}$; but this is clear since $(s_{t})$ is
generated, as a flow, by a quadratic Hamiltonian, and that the unitary
one-parameter group of operators determined by such a function precisely
consists of metaplectic operators.

\subsection{Proof: the general case}

We now no longer assume that $(f_{t})$ and $(F_{t})$ are one-parameter groups.
Recall that we have reduced the study of a time-dependent Hamiltonian
$H=H(z,t)$ to by introducing
\begin{equation}
\widetilde{H}(x,p,t,E)=H(x,p,t)-E
\end{equation}
which is a time-independent Hamiltonian on $\mathbb{R}^{2n+2}\equiv
\mathbb{R}_{x,p}^{2n}\times\mathbb{R}_{E}\times\mathbb{R}_{t}$ where $E$ is
viewed as conjugate variable to $t$. The flow $(\widetilde{f}_{t}^{H}%
)=(f_{t}^{\widetilde{H}})$ on $\mathbb{R}^{2n+2}$ generated by $\widetilde{H}$
is related to the time-dependent flow $(f_{t,t^{\prime}}^{H})$ by the formula
\begin{equation}
\widetilde{f}_{t}^{H}(z^{\prime},t^{\prime},E^{\prime})=(f_{t,t^{\prime}}%
^{H}(z^{\prime}),t+t^{\prime},E_{t,t^{\prime}})
\end{equation}
where $E_{t,t^{\prime}}-E^{\prime}$ is the variation of the energy in the time
interval $[t^{\prime},t]$. The advantage of this reformulation of the dynamics
associated with $H$ is that $(\widetilde{f}_{t}^{H})$ is a one-parameter group
of canonical transformations of $\mathbb{R}^{2n+2}$. In the operator case we
can proceed in a quite similar way, noting that the Weyl operator associated
with $\widetilde{H}$ is given by%
\[
\widehat{\widetilde{H}}=\widehat{H}(x,-i\hbar\nabla_{x},t)-i\hbar
\frac{\partial}{\partial t}.
\]
Of course $\widehat{\widetilde{H}}$ is self-adjoint if and only if
$\widehat{H}$ is, which is the case since $H$ is real. We will need the
following elementary fact, which is a variant of the method of separation of variables:

\begin{lemma}
Let $E$ be an arbitrary real number. The function
\begin{equation}
\Psi(x,t;t^{\prime})=\psi(x,t)e^{\frac{i}{\hbar}E(t-t^{\prime})} \label{psit}%
\end{equation}
is a solution of the extended Schr\"{o}dinger equation%
\begin{equation}
i\hbar\frac{\partial\Psi}{\partial t^{\prime}}=\widehat{\widetilde{H}}%
\Psi\label{extend}%
\end{equation}
if and only if $\psi=\psi(x,t)$ is a solution of the usual Schr\"{o}dinger
equation%
\begin{equation}
i\hbar\frac{\partial\psi}{\partial t}=\widehat{H}\psi. \label{classic}%
\end{equation}

\end{lemma}

\begin{proof}
We first note the obvious identity%
\begin{equation}
i\hbar\frac{\partial\Psi}{\partial t^{\prime}}=E\Psi. \label{id1}%
\end{equation}
Writing for short $\widehat{H}(t)=\widehat{H}(x,-i\hbar\nabla_{x},t)$ we have,
after a few calculations%
\begin{equation}
\left(  \widehat{H}(t)-i\hbar\frac{\partial}{\partial t}\right)  \Psi=\left[
\widehat{H}(t)\psi-i\hbar\frac{\partial\psi}{\partial t}\right]  e^{\frac
{i}{\hbar}E(t-t^{\prime})}+E\Psi\label{id2}%
\end{equation}
hence (\ref{extend}) is equivalent to (\ref{classic}) in view of (\ref{id1}).
\end{proof}

This result shows the following: choose an initial function $\psi_{0}=\psi
_{0}(x)$ at time $t=0$ and solve the usual Schr\"{o}dinger equation
(\ref{classic}), which yields the solution $\psi=\psi(x,t)$. Then $\Psi
=\Psi(x,t;t^{\prime})$ defined by (\ref{psit}) is the solution of the extended
Schr\"{o}dinger equation (\ref{extend}) with initial datum $\Psi
(x,t;t)=\psi(x,t)$ at time $t^{\prime}=t$. In terms of flows we can rewrite
this as
\[
\widetilde{F}_{t^{\prime}-t}(F_{t}\psi_{0})=(F_{t}\psi_{0})e^{\frac{i}{\hbar
}E(t-t^{\prime})}%
\]

\section{Conclusion}

In this paper we have shown how one can mathematically derive rigorously the
Schr\"{o}dinger equation from Hamiltonian mechanics. In that proof Banyaga's
theorem \cite{Banyaga} was seen to play a key role, when used in conjunction
with the Weyl formalism. In using the latter, it was necessary to introduce a
scaling factor which we arbitrarily chose to be $\hbar$ even though we made no
appeal to any quantum process. In this way we have shown that the mathematical
formalism of the theory of Schr\"{o}dinger's equation is already present in
classical mechanics, and is in fact a reformulation of Hamiltonian dynamics in
terms of operators. So where does quantum \textit{physics} enter the scene?
The most obvious question is, of course, how do we give a physical meaning to
the constant $\hbar$. That is: \emph{Why do we need Planck's constant?} Let us
explore a few possibilities; the list is certainly not exhaustive, and the
choice has been done in accordance with the present authors' tastes (and prejudices!).

\paragraph{Spectral properties of operators}

In a sense the most obvious and naive way to give a physical motivation for
the need of Planck's constant is, no doubt, empirical. We have seen that there
is a one-to-one correspondence between Hamiltonian flows and unitary evolution
operators solutions of Schr\"{o}dinger equations. Now, the spectral properties
of the involved Hamiltonian operators are well-understood; a basic postulate
of traditional quantum mechanics is that the eigenvalues of an operator are
the values that the corresponding observable can take. Using this postulate
one can thus put in the right value of the constant $\hbar$ by hand. This is
an empirical motivation, but it seems hard to refute because we know that it
works! It is irrefutable from an epistemological point of view: putting in
$\hbar$ that way we are sure to get the right physical answers!

\paragraph{The Narcowich--Wigner spectrum of a mixed state}

Consider a density matrix $\widehat{\rho}$: by definition it is a self adjoint
non-negative trace class operator with trace unity: $\widehat{\rho
}=\widehat{\rho}^{\ast}$, $\widehat{\rho}\geq0$, $\operatorname{Tr}%
\widehat{\rho}=1$. As is well-known from operator theory it is the positivity
property $\widehat{\rho}\geq0$ which is generally the most difficult to check;
one explicit tool is provided by the Kastler--Loupias--Miracle Sole (KLM)
conditions \cite{K,LM1,LM2}, but these imply the simultaneous verification of
infinitely many conditions (mathematically the KLM\ conditions are a
symplectic variant of Bochner's positivity criteria for the Fourier transform
of a measure). Narcowich \cite{NA} and Narcowich and O'Connell \cite{NACO}
have noticed the following property of mixed quantum states: the positivity
property depends in a crucial way on the numerical value of $\hbar$. That is,
if we change $\hbar$ some classical states become \textquotedblleft
quantum\textquotedblright\ (that is, representable by a density matrix), and
some quantum states loose this property (their density matrix is no longer
non-negative). The set of all values of $\hbar$ for which $\widehat{\rho}$
remains positive is then called the Narcowich--Wigner spectrum of
$\widehat{\rho}$.

\paragraph{Information theory}

Perhaps, after all, the answer ultimately lies in information theory. It might
very well be that the discrete nature of information is the key for the
passage from classical mechanics to quantum theory. Paraphrasing Anton
Zeilinger \cite{Zeilinger}:

\begin{quotation}
\textit{In conclusion it may very well be said that information is the
irreducible kernel from which everything else flows. The question why Nature
appears quantized is simply a consequence of the fact that information itself
is quantized. It might even be fair to observe that the concept that
information is fundamental is very old knowledge of humanity, witness for
example the beginning of gospel according to John: \textquotedblleft In the
beginning was the Word and the Word was with God, and the Word was
God\textquotedblright.}
\end{quotation}

\begin{acknowledgement}
The authors would like to thank the referees for valuable comments and
suggestions. We thank in particular the first reviewer for having pointed out
that condition (i) in Theorem \ref{TheoCo} cannot be relaxed, and for having
provided us with a counterexample.
\end{acknowledgement}


\begin{thebibliography}{99}                                                                                               %


\bibitem {AMR}Abraham, R., Marsden, J.E., Ratiu. T.: Manifolds, Tensor
Analysis, and Applications. Applied Mathematical Sciences \textbf{75},
Springer (1988)

\bibitem {Banyaga}Banyaga, A.: Sur la structure du groupe des
diff\'{e}omorphismes qui pr\'{e}servent une forme symplectique. Comm. Math.
Helv. \textbf{53}, 174--227 (1978)

\bibitem {banybook}Banyaga, A.: The structure of classical diffeomorphism
groups, Kluwer Academic Publishers (1997)

\bibitem {Derbes}Derbes, D.: Feynman's derivation of the Schr\"{o}dinger
equation. Amer. J. Math. Phys. \textbf{64}(7), 881--884, (1996)

\bibitem {Feynman2}Feynman, R.P.,:Space-time approach to non-relativistic
quantum mechanics. Rev. Modern Physics, \textbf{20}, 367--387 (1948)

\bibitem {Feynman}Feynman, R.P., Leighton, R.B., Sands, M.: The Feynman
Lectures on Physics, III, p.16-12, Addison-Wesley, Reading, MA (1965)

\bibitem {ICP}de Gosson, M.: The Principles of Newtonian and Quantum
Mechanics; with a Foreword by B. Hiley\textit{. }Imperial College Press,
London, 2001.

\bibitem {Birk}de Gosson, M.: Symplectic Geometry and Quantum Mechanics,
Birkh\"{a}user, Basel, (2006)

\bibitem {FP}de Gosson, M.: The Symplectic Camel and the Uncertainty
Principle: The Tip of an Iceberg? Found. Phys. \textbf{99,} 194--214 (2009)

\bibitem {PR}de Gosson, M., Luef, F.: Symplectic Capacities and the Geometry
of Uncertainty: the Irruption of Symplectic Topology in Classical and Quantum
Mechanics. Physics Reports \textbf{484,} 131--179 (2009)

\bibitem {Gro}Gr\"{o}chenig, K.: Foundations of Time-Frequency Analysis.
Birkh\"{a}user, Boston (2001)

\bibitem {Gromov}Gromov, M.: Pseudoholomorphic curves in symplectic
manifolds.\ Invent. Math. 82, 307--47 (1985)

\bibitem {GS}V. Guillemin, V., Sternberg, S.: Symplectic Techniques in
Physics. Cambridge University Press, Cambridge, Mass., 1984.

\bibitem {Hall}Hall M.J.W., Reginatto, M.: Schr\"{o}dinger equation from an
exact uncertainty principle. J. Phys. A: Math. Gen. \textbf{35} 3289 (2002)

\bibitem {K}Kastler, D.: The $C^{\ast}$-Algebras of a Free Boson Field.
Commun. math. Phys. \textbf{1}, 114--48 (1965)

\bibitem {lan}Lanczos, C.: The Variational Principles of Mechanics. Toronto,
Ontario: University of Toronto Press (1949). Reprint 4th edn. New York: Dover
Publications (1986)

\bibitem {Littlejohn}Littlejohn, R.G.: The semiclassical evolution of wave
packets. Physics Reports \textbf{138(}4--5), 193--291 (1986)

\bibitem {LM1}Loupias, G., Miracle-Sole, S.: $C^{\ast}$-Alg\`{e}bres des
syst\`{e}mes canoniques, I. Commun. math. Phys. \textbf{2}, 31--48 (1966)

\bibitem {LM2}Loupias, G., Miracle-Sole, S.: $C^{\ast}$-Alg\`{e}bres des
syst\`{e}mes canoniques, II. Ann. Inst. Henri Poincar\'{e} \textbf{6}(1),
39--58 (1967)

\bibitem {Mackey}Mackey, G.W.: The Relationship Between Classical and Quantum
Mechanics. In Contemporary Mathematics \textbf{214}, Amer. Math. Soc.,
Providence, RI (1998)

\bibitem {NA}Narcowich, F.J., O'Connell, R.F.: Necessary and sufficient
conditions for a phase-space function to be a Wigner distribution. Phys. Rev.
A \textbf{34}(1), 1--6 (1986)

\bibitem {NACO}Narcowich, F.J.: Geometry and uncertainty. J.\ Math. Phys.
\textbf{31}(2) (1990)

\bibitem {Nelson}Nelson, E.: Derivation of the Schr\"{o}dinger Equation from
Newtonian Mechanics. Phys. Rev. A, \textbf{150}(4):6, 1079--1085 (1966)

\bibitem {Polter}Polterovich, L.: The Geometry of the Group of Symplectic
Diffeomorphisms. Lectures in Mathematics, Birkh\"{a}user (2001)

\bibitem {RS}Reed M., Simon B.: Methods of Modern Mathematical Physics.
Academic Press, New York (1972)

\bibitem {schmelz1}Schmelzer, I.: Why the Hamiltonian Operator Alone Is not
Enough. Found. Phys. \textbf{39}(5), 486--498 (2009)

\bibitem {schmelz2}Schmelzer, I.: Pure Quantum Interpretations are not Viable.
Found. Phys. (2010)

\bibitem {es26a}Schr\"{o}dinger, E.: Quantisierung als Eigenwertproblem, Ann.
der Physik, \textbf{384}, (1926), 361-376.

\bibitem {Shubin}Shubin, M.A.: Pseudodifferential Operators and Spectral
Theory, Springer-Verlag (1987) [original Russian edition in Nauka, Moskva (1978)]

\bibitem {Stein}Stein, E.M.: Harmonic Analysis: Real Variable Methods,
Orthogonality, and Oscillatory Integrals. Princeton University Press (1993)

\bibitem {Stone}Stone, M.H.: Linear transformations in Hilbert space,
III:\ operational methods and group theory. Proc. Nat. Acad. Sci. U.S.A,
172--175 (1930)

\bibitem {struck}Struckmeier, J.: Hamiltonian dynamics on the symplectic
extended phase space for autonomous and non-autonomous systems. J. Phys. A:
Math. Gen. \textbf{38}, 1257--1278 (2005)

\bibitem {synge}Synge, J.L.: Encyclopedia of Physics, vol \textbf{3/1} ed. S.
Fl\"{u}gge, Berlin: Springer (1960)

\bibitem {Wong}Wong, M.W.: Weyl Transforms, Springer-Verlag (1998).

\bibitem {Zeilinger}Zeilinger, A.: http://www.metanexus.org/ultimate\_reality/zeilinger.pdf
\end{thebibliography}
\end{document}